\documentclass[aps,twocolumn,english]{revtex4}

\usepackage[T1]{fontenc}
\usepackage[latin9]{inputenc}
\usepackage{textcomp}
\usepackage{amsmath}
\usepackage{graphicx}
\usepackage{amssymb}
\usepackage{babel}

\begin{document}

\title{Subnanosecond spectral diffusion measurement using photon correlation}

\author{G. Sallen$^{1}$, A. Tribu$^{2}$, T. Aichele$^{1,3}$, R. Andr\'{e}$^{1}$, L. Besombes$^{1}$, C. Bougerol$^{1}$, M. Richard$^{1}$, S. Tatarenko$^{1}$, K. Kheng$^{2}$, and J.-Ph.~Poizat$^{1,*}$}

\affiliation{$^1$ CEA-CNRS-UJF group 'Nanophysique et Semiconducteurs',
 Institut N\'{e}el, CNRS - Universit\'{e} Joseph Fourier, 38042 Grenoble, France, \\
$^2$ CEA-CNRS-UJF group 'Nanophysique et Semiconducteurs', CEA/INAC/SP2M, 38054 Grenoble, France, \\
$^3$Physics Institute, Humboldt University, Berlin, Germany, \\
* email : jean-philippe.poizat@grenoble.cnrs.fr}



\maketitle


\textbf{Spectral diffusion  corresponds to random spectral jumps of a narrow line as a result of fluctuating environment.
It is a prominent issue in spectroscopy since the observed spectral broadening prevents access to intrinsic line properties. On the other hand, its characteristic  parameters provide  local information on the environment of a light emitter embedded in a solid matrix, or moving within a fluid, with numerous applications in physics or biology.
We present a new experimental technique for measuring spectral diffusion
based on photon correlations  within a spectral line.
 Autocorrelation on  half of the line as well as cross-correlation between the two halves give a quantitative value of the spectral diffusion time with a resolution only limited by  the correlation set-up.
 We have measured spectral diffusion of the photoluminescence of a single light emitter
 with  $90$ ps time resolution, exceeding by  four orders of magnitude the best reported  resolution.
}

Spectral diffusion (SD) was first studied in spin resonance experiments \cite{Klauder} and observed since then in various light emitting systems such as rare-earth ions \cite{Flach}, ruby \cite{Szabo}, molecules \cite{Ambrose,Zumbusch} or semiconductor quantum dots \cite{Empedocles,Robinson,Turck,Besombes,Empedocles97}.
SD of a single emitter results from its fluctuating  environment \cite{Ambrose,Zumbusch,Empedocles,Robinson,Turck,Besombes,Empedocles97}. It is generally due to Stark effect caused by  randomly trapped charges in the vicinity of the emitter \cite{Empedocles97}.
In this work we are interested in light emitting semiconducting nanostructures and more specifically in quantum dots (QDs). Such nanostructures are very promising objects for quantum information or laser physics.
Understanding their luminescence linewidth is obviously of primary importance in either of these applications where such an emitter is to be coupled to another emitter or to an optical cavity. Optical coherent control of a single qubit encoded on the spin of a QD can be implemented only in the absence of SD. The SD characteristic time gives the maximum time under which the system can be considered as SD-free. More generally, if one wants to reduce  this  diffusion, it requires a good understanding of its origin and therefore an accurate measurement of its temporal behaviour.

The usual method to evidence SD of a single emitter is to record a time series of spectra and to visualize directly the spectral wandering \cite{Ambrose,Empedocles,Robinson,Turck,Besombes}.
The time resolution of this method is limited by the ability of acquiring a spectrum in a short time.
The counting rate from a single emitter can hardly exceed $10^5$s$^{-1}$ and it is therefore impossible to extract a spectrum on a time scale shorter than $10^{-5}$ seconds. In practise, single photon counting charged coupled devices (CCD) can not supply more than $1000$ images per second, so the time resolution can not be better than a few ms.

Better SD time resolution has been obtained on inhomogeneously broadened ensembles of semiconducting nanocrystals by measuring a modulation frequency dependent linewidth in a spectral hole burning experiment \cite{Palinginis}.
 With this technique the time resolution is set by the modulation frequency and Palinginis \emph{et al} \cite{Palinginis} have obtained a resolution in the $100 \mu s$ range. Another technique relying also on resonant absorption was developed by Zumbusch et al \cite{Zumbusch} for a single emitter.
 The absorption  of a narrow laser line is fluctuating owing to the spectral fluctuations of the emitter line via which the resonant excitation is performed.  The emitted fluorescence light undergoes therefore intensity fluctuations that are then measured by photon correlation. In their work, Zumbusch et al \cite{Zumbusch} report an ideal time resolution of $200 ns $ but their integration time sets a practical limit of  $1 \mu s$.

Reaching short time resolution with individual emitters has been achieved recently by using
photon-correlation Fourier spectroscopy (PCFS) \cite{Brokmann,Coolen1}. This  technique is based on  the intensity correlations at the two outputs of a Michelson interferometer. The fringe patterns at the two outputs are complementary for a given wavelength.
Spectral jumps can cause a bright fringe  to become  dark on one output and appear bright on the other. This leads to changes in the output intensities on a characteristic time given by the SD time.
The theoretical time resolution  is given by the photon correlation set-up as in the method we present here.
However PCFS requires interferometric stability and
Coolen \emph{et al} \cite{Coolen} have only reached a time resolution of  $20 \mu s$  owing to drift problems in their set-up.

\begin{figure*}[t]
\resizebox{0.9\textwidth}{!}{\includegraphics{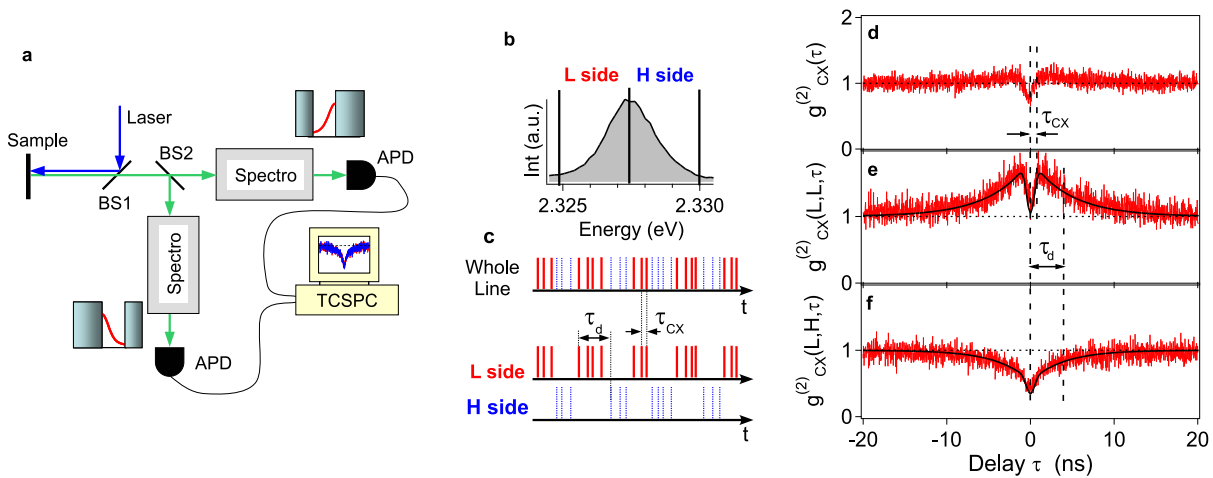}}
 \caption{\textbf{Spectral diffusion measurements with photon correlations.}
  (a), Experimental set-up. BS1 and BS2 are beamsplitters with  transmission T1=70\% and T2=50\% respectively. TCSPC is for Time Correlated Single Photon Counting data acquisition card. APD is for Avalanche Photodiode.
   (b), Photoluminescence spectrum of the charged exciton transition of a single QD integrated during 1 s. The two halves of the line are labeled L and H.
(c), Representation  of the photon time distribution in the two halves of the profile
 (d), Auto-correlation of the whole profile.
 (e), Auto-correlation of one half of the profile showing the bunching due to SD ($\tau_d=4$ ns) and the narrower single photon antibunching.
 (f), Cross-correlation between the two halves of the profile displaying the antibunching due to SD with the same characteristic time $\tau_d=4$ ns as above.
 All data (d-f) shown in this figure have been obtained on the same QD with the same excitation power.
 In (e,f) the solid lines are fits with the model explained in the text and the Supplementary information.
}
 \label{fig:SD}
\end{figure*}

In this letter, we present  a new and simple photon correlation technique to access characteristic SD times of a single emitter with a subnanosecond resolution. This is, to our knowledge, four orders of magnitude better than the best time resolution so far \cite{Zumbusch,Coolen}. Our technique is very robust since it is phase-insensitive and relies on linear optics. It is based on correlations of photons emitted within a spectral window narrower than the SD broadened inhomogeneous line
  (cf fig. \ref{fig:SD}(a-c)). Owing to the wandering of the homogeneous line, the emission peak stays a limited time within this spectral window leading
to photon bunching with the characteristic time $\tau_d$ on the autocorrelation on one half of the line  (see fig. \ref{fig:SD}(e)), and to photon antibunching with the same characteristic time $\tau_d$ for the cross-correlation between the two halves of the profile (see fig. \ref{fig:SD}(f)).
 The resolution is then only limited by  the photon correlation set-up (see Methods).
 The minimum time delay between photons is of the order of $\tau_{CX}=600ps$ and leads to the narrow antibunching dip in (d) and (e).

We have applied this technique  to individual semiconducting quantum dots (QDs) embedded in a nanowire (NW). Just like semi-conducting nanocrystals \cite{Empedocles, Empedocles97, Palinginis, Coolen}, these QDs suffer from ultrafast fluctuations caused by the vicinity of  surface states, as opposed to usual encapsulated self-assembled QDs.
Details on the growth of the CdSe/ZnSe NWs can be found in  \cite{Aichele}.
Exciton (X), biexciton (XX) and charged exciton (CX) transitions have been identified unambiguously using photon correlation spectroscopy \cite{correlation}.  The radiative lifetimes of these transitions  are respectively $\tau_X=700$ ps, $\tau_{XX}=400$ ps, $\tau_{CX}=600$ ps. The luminescence wavelength is around $550$ nm with a high count rate of $25 000$ counts per second at $T=4$K.
This system has demonstrated single photon operation up to a temperature of $220$ K \cite{Tribu}.
The microphotoluminescence experimental set-up is described in the Methods section.

In this work we have been focusing first on the CX line shown in fig \ref{fig:SD}(b).  A series of spectra taken every $0.15$ s during $25$ s does not exhibit any visible SD. The lineshape is better fitted with a Gaussian than with a Lorentzian. A Gaussian shape is characteristic for an inhomogeneous broadening mechanism like SD  \cite{Berthelot}.
In fig. \ref{fig:SD}(d), the autocorrelation of the whole profile exhibits the characteristics antibunching  dip  of a single photon source. It is not very pronounced since the timing resolution of the experimental set-up ($800$ ps) is of the same order as the emitter's lifetime ($\tau_{CX}=600$ ps). The slight bunching is due to the hopping between the neutral and the charged state of the QD. All these features have been discussed in detail on data obtained with a $90$ ps resolution  set-up \cite{correlation}.
Fig. \ref{fig:SD}(e)  displays the result of auto-correlation on one half of the emission peak. In addition to the clear antibunching dip (width $\sim\tau_{CX}$) characteristic of single photon emission, it exhibits a clear bunching feature that  is significantly larger than that obtained with the whole line profile in fig. \ref{fig:SD}(d).
This demonstrates that the homogeneous line remains during a characteristic time $\tau_d=4$ ns within one half of the SD broadened line as illustrated in fig. \ref{fig:SD}(c).
Fig \ref{fig:SD}(f) shows cross-correlation measurements between the low and high energy sides of the emission spectrum. It exhibits a broad antibunching on the same time scale $\tau_d$ as the bunching peak of fig \ref{fig:SD}(e). This
is different from the single photon antibunching and is a clear signature of SD showing that it takes a characteristic time $\tau_d$ for the intrinsic line to move from one half of the spectral profile to the other (cf  fig \ref{fig:SD}(c)).

Autocorrelation and cross-correlation methods yield the same  SD time $\tau_d$ for the emitter regardless of the width and position of the non-overlapping spectral windows, provided that the widths and energy difference of the latter are larger than the intrinsic homogeneous linewidth of the wandering line, which is generally the case.
This makes this technique very robust. We have checked this non-trivial property experimentally (cf  fig \ref{fig:Spectral-bins}), and a theoretical analysis is given in the Supplementary Information.
 This property can be understood by considering the fact that
 the characteristic time of photon correlation measurements is given by short time events.

 In autocorrelation experiments, the homogeneous line does not move for time scales shorter than $\tau_d$. After a time larger than $\tau_d$ the memory of the spectral position is lost so the number of autocorrelation events drops as can be seen in fig. \ref{fig:SD}(e). As demonstrated in the Supplementary Information, the width of the bunching peak does not depend on the size of the spectral window; only the height of the peak does.

 In a similar manner, the cross-correlation function exhibits a lack of events for $|\tau | < \tau_d$ (antibunching dip in fig. \ref{fig:SD}(f)) corresponding to situations where the homogeneous line has not had time to  hop to a  position within the other spectral window.
 As shown in fig. \ref{fig:Spectral-bins}(a), we have  experimentally checked that the diffusion time $\tau_d$ extracted from the data does not depend on the widths of the two spectral windows, as derived in the Supplementary Information in the case of spectral windows larger than the homogeneous linewidth.

We also observed in fig. \ref{fig:Spectral-bins}(d) that the energy separation between the two spectral windows in the cross-correlation set-up has no influence. This is due to the fact that, at each spectral jump, the new position is independent of the former and is randomly distributed over the whole inhomogeneous profile with a Gaussian probability.

\begin{figure}
\vspace{2cm}
\resizebox{0.45\textwidth}{!}{\includegraphics{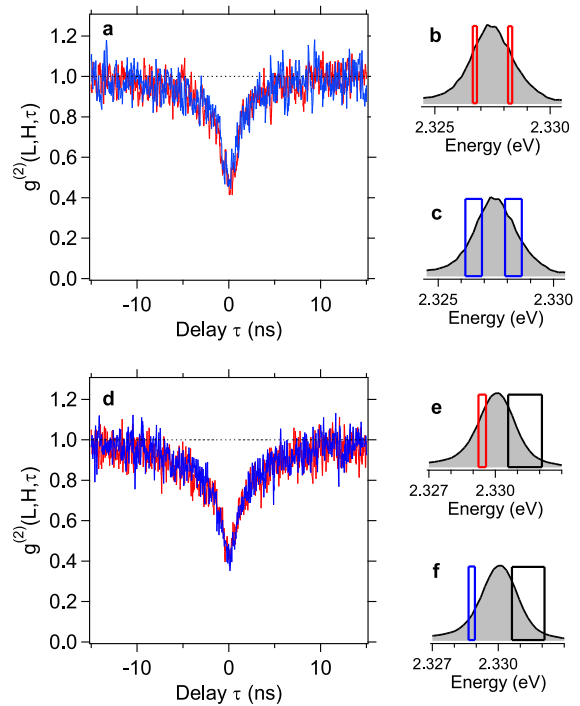}}
 \caption{ \textbf{Influence of the width (a-c) and position (d-f) of the spectral windows in cross correlations.}  (a), The red (resp. blue) trace is the cross correlation between spectral windows    of width $0.3$meV (resp. $1.0$meV), as depicted in (b) (resp.(c)). (d),  Cross correlation between spectral windows with different energy separations. The red (resp. blue) trace  corresponds to the situation depicted in (e) (resp.(f)).
 In (a) and in (d), the red and blue traces are superimposed  and yield therefore the same characteristic time $\tau_d$.
 Note that the pumping power for (a) is higher ($15\mu$W) than in (d) ($10\mu$W). These different experimental conditions are the reason for the different  $\tau_d$ value between (a)  and (d). }
 \label{fig:Spectral-bins}
 \end{figure}

\begin{figure}
 \resizebox{0.3\textwidth}{!}{\includegraphics{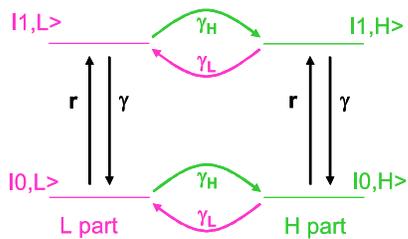}}
 \caption{  \textbf{Simplified level scheme used for the rate equation model of SD.} The QD emits photons either on the low  (L) or high (H) energy side of the emission peak. The QD emission energy diffuses with a rate $\gamma_H$ (resp $\gamma_L$) from the $L$ (resp. $H$) to the $H$ (resp. $L$) part. On each energy side, the QD is modeled with two levels   containing respectively zero, or one exciton. Note that  other levels are included for the fits (see text). The quantities $r$ and $\gamma$  are respectively the incoherent pumping rate and the exciton decay rate.    }
 \label{fig:model}
\end{figure}

The diffusion time $\tau_d$ is extracted by fitting experimental data, and its value is more accurate if $\tau_d$ is of same order or larger than the lifetime of the emitter.
The principle of the  model is depicted in fig. \ref{fig:model} in the simplified case of  two complementary spectral windows. Each spectral window is modeled as a two-level system.
Each level is connected  to its counterpart with a "jump" rate $\gamma_i$ ($i=L,H$). As shown in the Supplementary Information, the SD rate is given by $\gamma_d=\sum_i \gamma_i$ and is extracted directly from the time-width of the correlation measurement.
In practice, the fitting model allows for non-complementary spectral windows,  and uses a multi-level system to account for   the neutral (X) and charged (CX) excitons,   dark exciton, and neutral biexciton (see  Supplementary Information and \cite{correlation}).

\begin{figure}
 \resizebox{0.45\textwidth}{!}{\includegraphics{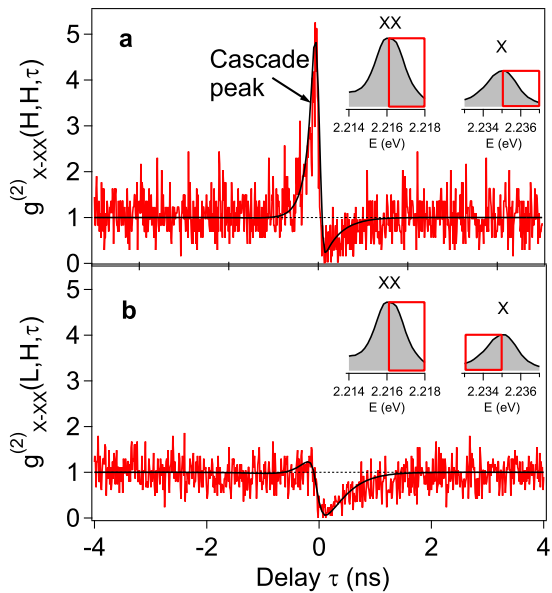}}
 \caption{\textbf{Correlated spectral diffusion between two lines}.  (a), Cross-correlation between the high energy sides (H) of the exciton (X) and the biexciton (XX) (see inset) exhibiting the characteristic biexciton-exciton cascade peak.
(b), Cross-correlation between the low energy sides (L) of the exciton (X) and the high energy side (H) of the biexciton (XX) lines (see inset). The absence of the characteristic cascade peak shows that  two photons originating from different halves are not part of the same cascade.
For (a,b), the solid line is a fit based on the model depicted in fig. \ref{fig:model} of the letter where extra levels have been  added as done in \cite{correlation}.}
 \label{fig:X-XX}
\end{figure}

An extra feature of this photon correlation technique is that it allows the investigation of correlated spectral diffusion between different lines  coming either from the same emitter or from different emitters.
This possibility gives information on the energy shift of different lines caused by  a common  change in their environment.
We have  performed cross-correlation between one half of the exciton (X) line and one half of the biexciton (XX) line of the same quantum dot.
This experiment has been performed with fast APDs leading to a correlation timing resolution of $90$ ps (see Methods).
The results are shown in fig. \ref{fig:X-XX}  for the H half of the X line correlated either with the H half of XX line (fig. \ref{fig:X-XX}(a)) or with its L half (fig. \ref{fig:X-XX}(b)). The characteristic cross-correlation peak of the biexciton-exciton cascade \cite{Moreau} can clearly be seen in fig. \ref{fig:X-XX}(a) for the same halves of the profile, whereas it is missing in fig. \ref{fig:X-XX}(b). This shows clearly that the sign of the energy shifts of the X and XX lines due to the fluctuating environment is the same, as already observed in \cite{Besombes} at slower time scales.

To summarize, we have presented a simple and robust method to measure spectral diffusion of single emitters with a resolution of $90$ ps. As shown experimentally, this technique enables the probing of the fluctuations of the nanoenvironment of a single emitter with a time resolution improved by four orders of magnitude compared to existing achievements.

\section*{Methods}

\textbf{Experimental set-up.} It is based on a standard microphotoluminescence experiment operating at a temperature of T=4K.
The sample is excited by a continuous wave diode laser emitting at $405$ nm.
The luminescence  is  split by a 50/50 beamsplitter and each beam is sent to a monochromator (resolution $\delta E=0.2$ meV or $\delta\lambda=0.05$ nm) whose output slit is imaged on an avalanche photodiode (APD).
The width of the output slit can be varied allowing us to choose the spectral window within the SD  inhomogeneously broadened line.
The voltage pulses of each APD are sent to
a time-correlated single photon  module that
builds an histogram of the time delays between  photons.
This allows us to perform either auto-correlation when the two monochromators are tuned to the same wavelength or cross-correlation otherwise.
Except for the results of fig. \ref{fig:X-XX}, the work presented here has been obtained with high quantum efficiency APDs  ($\eta=60 \%$ at $550$ nm). High detection efficiency is important when performing correlation experiments since the  integration time is proportional to $\eta^2$. The price to pay is their slower timing resolution.
With these APDs, the measured timing resolution of the whole set-up is $800$ ps (full width at half maximum).
This rather slow time resolution is not a limitation in our case since the SD times that we are investigating are in the 10 ns range.
The results shown in fig. \ref{fig:X-XX} have been obtained with fast APDs to allow the observation of the temporally narrow structures. In that case the measured time  resolution of the set-up is $90$ ps, and the APD quantum efficiency is ($\eta=30 \%$).

\section*{Acknowledgments}

We acknowledge very efficient technical support of F. Donatini and careful reading of the manuscript by Le Si Dang and G. Nogues.
T.A. acknowledges support by Deutscher
Akademischer Austauschdienst (DAAD). Part of this
work was supported by European project QAP (Contract No.
15848).

\section*{Author contributions}

GS conducted the optical experiments and analysed the data. AT, TA, RA,  ST and KK took care of the fabrication and processing of the samples, and CB performed their structural analysis. GS, LB, MR and JPP contributed to the genesis of the idea and to the discussion of the results. JPP supervised the optical experiments and wrote the paper.

\newpage

\section*{Supplementary Information : Analysis of the influence of the size of the spectral windows}

In this supplementary information, we discuss theoretically the effect on correlation functions of the fact that the spectral detection windows defined by the two spectrometers are not exactly complementary and may have different sizes. This corresponds namely to the two situations depicted in Fig. S1: (a) the  two windows do not cover the whole line and do not overlap, (b)  the two windows cover the whole line and overlap.
In the  sections A and B below, we will consider the  common situation in which the sizes of the spectral windows are larger than the intrinsic homogeneous linewidth of the wandering line. In section C, we will address the case in which the spectral windows are narrower than the homogeneous linewidth.

\subsection{Rate equations in the large spectral window limit}
\label{sec:rate}

\begin{figure}[h]
\resizebox{0.25\textwidth}{!}{\includegraphics{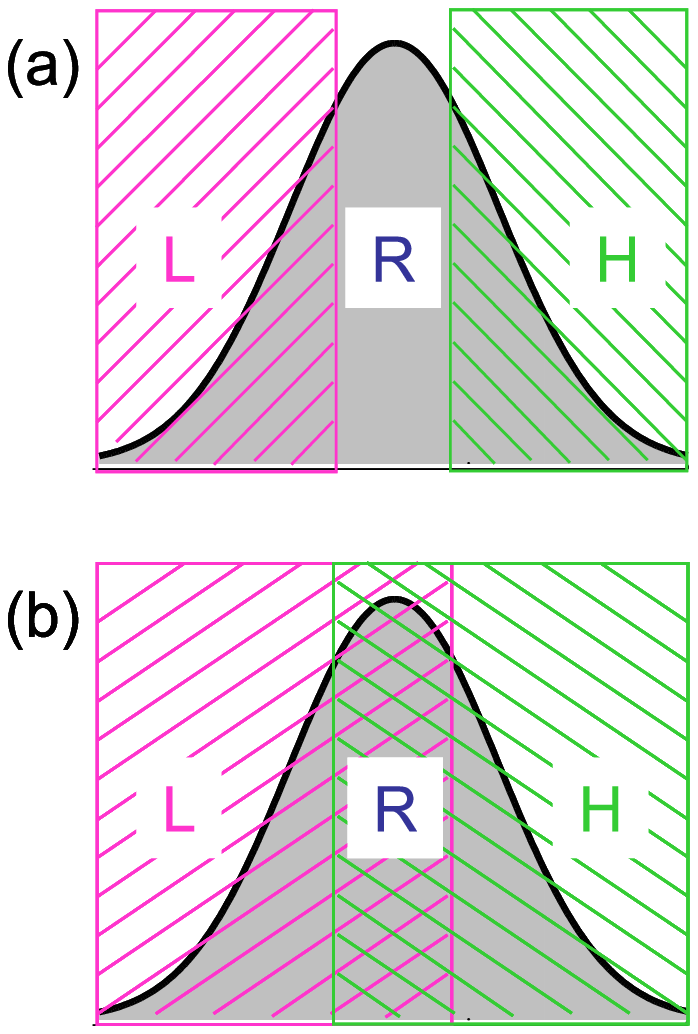}}
 \resizebox{0.4\textwidth}{!}{\includegraphics{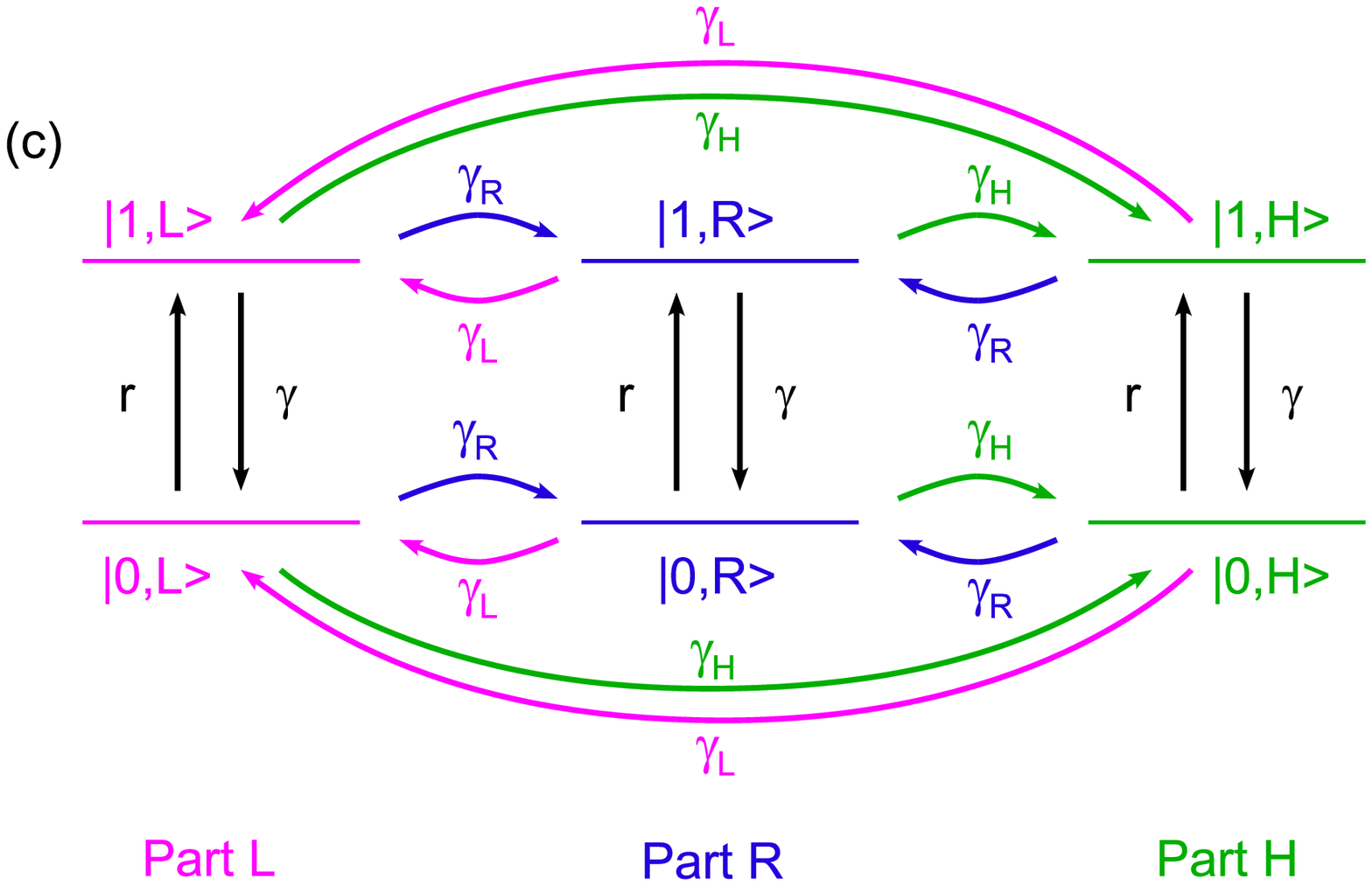}}
 \caption{  (a) Situation corresponding to two non-overlapping  windows that do not cover the whole line. The central region R is not detected by any of the spectrometers. (b) Situation corresponding to two overlapping spectral windows that cover the whole line. The central part R corresponds here to the overlap region. In this case, the spectral window covered by the first (resp. second) spectrometer is L+R (resp. H+R).
(c) Level scheme used for the rate equation model (see text).}
 \label{fig:suppl}
\end{figure}

The scheme shown in Fig. \ref{fig:suppl} involves three parts corresponding to different spectral regions. Each part consists in  a two-level system with a ground and an excited state connected by the pumping rate $r$ and the spontaneous emission rate $\gamma$.
Each part diffuses to the other two parts with a rate $\gamma_i$, ($i=L,H,R$)  depending on the  part to which the diffusion occurs.
In this section, we assume that the sizes of the different parts are larger than the homogeneous linewidth.

The six rate equations corresponding to the scheme shown in Fig. \ref{fig:suppl}(c) can be written in the following form after straight forward algebra :
\begin{eqnarray}
 \frac{d n_{0i}}{dt} &=& -r n_{0i} + \gamma n_{1i} - \gamma_d n_{0i} + \gamma_i N_0  , \nonumber \\
 \frac{d n_{1i}}{dt} &=& -\gamma n_{1i} + r n_{0i} - \gamma_d n_{1i} + \gamma_i N_1 ,
 \label{eq:rate}
 \end{eqnarray}
where $i= L,H,R$. The quantity $n_{0i}$ (resp. $n_{1i}$) is the population of the level $|0,i\rangle$ (resp. $|1,i\rangle$). We define $N_0=\sum_i n_{0i}$ and $N_1=\sum_i n_{1i}$, with $N_0 + N_1=1$. The quantity
$\gamma_i$ is the rate towards spectral window $i$.
 We also define $\gamma_d=\sum_i \gamma_i$.
 The individual rate $\gamma_i$, ($i=L,H,R$) is proportional to the line area defined by part $i$ and  fulfills  $\gamma_i /\gamma_d= N_i(t=\infty)$. It depends therefore on the window size and position. But, as it will appear below, only the sum  $\gamma_d=\sum_i \gamma_i$ shows up in the time scale involved in the expressions of the correlation functions.
 The quantity $\gamma_d$ turns out to be the intrinsic rate corresponding to the spectral diffusion of the light emitter under consideration.

Straight forward algebra  leads to the following simple differential equation for $N_0$ and $N_1$ :
\begin{equation}
\frac{d N_{1}}{dt} = -\gamma N_1 + r N_0.
\end{equation}
Its solution is given by
\begin{equation}
N_1 (t)= \frac{r}{r+\gamma} + C \exp(-(r+\gamma)t),
\end{equation}
where $C$ is a constant depending on the initial conditions.

Let us focus now on the total population $N_i=n_{0i}+n_{1i}$ of window $i$ ($i=L,H,R$). We obtain
\begin{equation}
\frac{d N_{i}}{dt} = -\gamma_d N_i + \gamma_i.
\end{equation}
 The solution of this equation is given by
\begin{equation}
N_i (t)= \frac{\gamma_i}{\gamma_d} + K \exp(-\gamma_d t),
\end{equation}
where $K$ is a constant depending on initial conditions.
It can be noticed that the only time scale describing the population evolution $N_i(t)$ of part $i$ is the total spectral diffusion rate $\gamma_d$.

It can easily be shown that the general solution to equations \ref{eq:rate} for $n_{0i}$, and $n_{1i}$ is
\begin{eqnarray}
n_{0i}(t)=N_0(t)N_i(t),  \nonumber \\
n_{1i}(t)=N_1(t)N_i(t),
\label{eq:n=NN}
\end{eqnarray}

\subsection{Derivation of the correlation functions in the large spectral window limit}

\subsubsection{Non-overlapping spectral windows}

We discuss here the situation corresponding to Fig. \ref{fig:suppl}(a). We will show here that the size of the non-overlapping spectral windows do not affect the  spectral diffusion time $\tau_d$ measured by photon correlation. These theoretical results support the experimental observation shown in the main paper (Fig. 2).

\vspace{2mm}

\paragraph{\textbf{Auto-correlation}}

Let us derive the autocorrelation function on the $L$ window.
 The intensity correlation function is given by $g^{(2)}(L,L,\tau)=n_{1L}(\tau)/n_{1L}(\infty)$, with the initial condition  $n_{0L}(0)=1$ corresponding to  the detection of a photon in this window. Using the results of the preceding section \ref{sec:rate}, we obtain
\begin{equation}
\begin{split}
 g^{(2)}(L,L,\tau)= & \left[1+\left(\frac{\gamma_d}{\gamma_L}-1\right)\exp(-\gamma_d \tau)\right]  \\
& \times \left[1-\exp(-(r+\gamma)\tau)\right].
\label{eq:nonoverlap-auto}
\end{split}
\end{equation}
The autocorrelation function is the product of a bunching envelope due to spectral diffusion (first term) and  the usual
 antibunching shape due to the single photon nature of the source (second term). The bunching  envelope due to the spectral diffusion has a width given by the spectral diffusion time $\tau_d=1/\gamma_d$ independently of the size of spectral window. As already mentioned the size of  window $L$ affects the rate $\gamma_L$ into this window and therefore only the height
 $(\gamma_d - \gamma_L)/\gamma_L$ of the bunching peak. In the limit where the window $L$ covers the whole spectral line ($ \gamma_L \rightarrow \gamma_d$), the spectral diffusion bunching peak disappears. In the case of a narrow spectral window ($\gamma_L \rightarrow 0$) the bunching peak becomes very high and scales as $\gamma_d/\gamma_L$.

\vspace{2mm}

\paragraph{\textbf{Cross-correlation}}

Let us derive now the cross-correlation between windows $L$ and $H$. We assume that the first photon is detected in window $L$, so the intensity cross-correlation is given by $g^{(2)}(L,H,\tau)=n_{1H}(\tau)/n_{1H}(\infty)$, with  the initial condition  $n_{0L}(0)=1$. Using the results of section \ref{sec:rate}, we obtain
\begin{equation}
g^{(2)}(L,H,\tau)=\left[1-\exp(-\gamma_d \tau)\right]\left[1-\exp(-(r+\gamma)\tau)\right].
\label{eq:nonoverlap-cross}
\end{equation}
The cross-correlation function is the product of a broad antibunching function corresponding to spectral diffusion and, as for the auto-correlation, the usual antibunching term due to the single photon nature of the source.
The expression of $g^{(2)}(L,H,\tau)$ shows that the cross-correlation does not depend on the sizes of the spectral windows.

\subsubsection{Overlapping spectral windows}


\paragraph{\textbf{Auto-correlation}}

We consider here the situation where the auto-correlation is performed with two spectrometers whose spectral windows might not match perfectly. This means that the window of the first spectrometer corresponds to the $L$ part and the window of the second spectrometer the $L$ and $R$ parts.

The intensity auto-correlation corresponds to a first detection in part $L$ and a second detection in $L$ or $R$ parts. It is given by
$g^{(2)}(L,L+R,\tau)=(n_{1L}(\tau)+n_{1R}(\tau))/(n_{1L}(\infty)+n_{1R}(\infty))$, with  the initial condition  $n_{0L}(0)=1$.
Using the results of eqs. \ref{eq:n=NN} of section \ref{sec:rate}, we can write
\begin{equation}
n_{1L}(\tau)+n_{1R}(\tau)= N_1(\tau)[N_L(\tau)+N_R(\tau)]
\end{equation}
Taking into account the initial conditions, we have
\begin{eqnarray}
N_1(\tau) &=& \frac{r}{r+\gamma}\left[1-\exp(-(r+\gamma)\tau )\right], \\
N_L(\tau) &=& \frac{\gamma_L}{\gamma_d}
\left[1+\left(\frac{\gamma_d }{ \gamma_L}-1 \right) \exp(-\gamma_d \tau)\right] , \\
N_R(\tau) &=&  \frac{\gamma_R}{\gamma_d}\left[1-\exp(-\gamma_d\tau)\right].
\end{eqnarray}
The auto-correlation function writes then
\begin{equation}
\begin{split}
g^{(2)}(L,L+R,\tau) = & \left[ 1+ \left(\frac{\gamma_d}{\gamma_R+\gamma_L}-1\right) \exp(-\gamma_d\tau)\right] \\
& \times \left[1-\exp(-(r+\gamma)\tau )\right].
\end{split}
\end{equation}
This expression is very close to eq. (\ref{eq:nonoverlap-auto}). Owing to the imperfect matching between the two spectrometer windows, the height of the spectral diffusion bunching peak is reduced to $\gamma_d/(\gamma_R+\gamma_L)-1$ compared to  $\gamma_d/\gamma_L-1$ in eq. (\ref{eq:nonoverlap-auto})).

\vspace{2mm}

\paragraph{\textbf{Cross-correlation}}

In the case of overlapping spectral windows (cf Fig. \ref{fig:suppl}(b)), the spectral region $R$ corresponds to the overlap between the spectral windows of the two spectrometers. This means that the  window of the first spectrometer covers the $L$ and $R$ parts, whereas the window of the other spectrometer covers the $R$ and $H$ parts. Let us mention that in this case the whole line is covered.

The intensity cross-correlation corresponds to a first detection in regions $L$ or $R$, and a second detection in regions $R$ and $H$. It is  given by $g^{(2)}(L+R,R+H,\tau)=(n_{1R}(\tau)+n_{1H}(\tau))/(n_{1R}(\infty)+n_{1H}(\infty))$, with  the initial condition  $n_{0L}(0)+n_{0R}(0)=1$. Using the results of eqs. \ref{eq:n=NN} of section \ref{sec:rate}, we have
\begin{equation}
n_{1R}(\tau)+n_{1H}(\tau)= N_1(\tau)[N_R(\tau)+N_H(\tau)]
\end{equation}
Taking into account the initial conditions, we have
\begin{eqnarray}
N_1(\tau) &=& \frac{r}{r+\gamma}\left(1-\exp(-(r+\gamma)\tau )\right), \\
 N_R(\tau) &=&  \frac{\gamma_R}{\gamma_d}\left[1+ \left( \frac{\gamma_d}{\gamma_R+\gamma_L}-1\right) \exp(-\gamma_d\tau )\right], \\
  N_H(\tau) &=&  \frac{\gamma_H}{\gamma_d}\left[1-\exp(-\gamma_d\tau)\right].
\end{eqnarray}
This leads to
\begin{equation}
\begin{split}
& g^{(2)}(L+R,H+R,\tau) = \\
&  \left[1-\frac{\gamma_L }{(\gamma_L +\gamma_R)} \frac{\gamma_H}{(\gamma_H +\gamma_R)}\exp(-\gamma_d\tau)\right] \\
& \times \left[1-\exp(-(r+\gamma)\tau)\right] .
\end{split}
\end{equation}
This expression is to be compared with eq. (\ref{eq:nonoverlap-cross}).
The first term corresponds  to the antibunching caused by the spectral diffusion with a characteristic time $\tau_d$.
Owing to the overlap of the two spectral windows, this antibunching contrast is  reduced. This contrast is unity when the overlap is vanishing as in eq. (\ref{eq:nonoverlap-cross}). On the other hand, when the overlap is maximum, this correlation term is unity, and one is left with a the "single photon type" antibunching.

\subsection{Limit of vanishingly narrow spectral windows : determination of the intrinsic linewidth}
\label{sec:intrinsic}

In this section we discuss qualitatively the situation in which
the  spectral windows become vanishingly narrow, and  eventually reach a value smaller than the intrinsic homogeneous linewidth. We will show that in this very specific situation, the cross-correlation does depend on the size of the spectral windows and their energy difference.

Let us  consider the cross-correlation experiment  with both spectral windows narrower than the homogeneous linewidth. When the energy difference between the two spectral windows is larger than the homogeneous linewidth, we are in the situation described in paragraph B.1.b of this supplementary information, with the superposition of the antibunching due to the single photon nature of the source (time scale $1/(r+\gamma )$) and the generally broader antibunching due to the spectral diffusion (time scale $1/\gamma_d$). When the energy difference between the two spectral windows is reduced and becomes  smaller than the homogeneous linewidth, the homogeneous line will either include both spectral windows or none of them.
The contribution of the "spectral diffusion" type antibunching will then vanish and only the "single photon" type antibunching will remain. The energy difference corresponding to the transition between these two regimes will give the homogeneous linewidth.

So our method allows in principle the determination of the intrinsic homogeneous linewidth of the wandering line. In practise however, this experiment is very difficult to achieve for two reasons : i)  it requires a spectrometer with a resolution better than the homogeneous linewidth, ii) the total number of counts through narrow spectral windows will be very low, requiring very large integration time to acquire a meaningful photon correlation histogram.

Note that the photon correlation Fourier spectroscopy (PCFS) method of refs [12,13] of the main text is better suited for extracting the homogenous linewidth of a narrow wandering line. In this method the two spectral windows are complementary fringe patterns obtained at the two outputs of a Michelson interferometer. The fringe spectral spacing can be made arbitrarily small by increasing the optical path difference between the two arms, so that the resolution can be made arbitrarily high and exceed the homogeneous linewidth (these are the same reasons for which Fourier transform spectroscopy outperforms conventional spectroscopy with respect to resolution).
When the fringe spectral periodicity becomes narrower than the homogeneous linewidth (ie the optical path difference is large) the intensity noise is no longer  related to spectral diffusion. It is therefore possible to identify this transition and evaluate the homogeneous linewidth.
Since in PCFS the two spectral windows are complementary whatever the resolution, all the photons are always detected and contribute to the photon correlation histogram so that high resolution does not imply long integration time.
The PCFS method offers therefore a real possibility of extracting the intrinsic homogeneous linewidth.
The price to pay is the drastic mechanical stability requirement necessary to implement such interferometric experiments.

\end{document}